\documentclass[a4paper]{jpconf}
\usepackage{graphicx}
\usepackage{subfig}
\usepackage{slashed}
\usepackage{mathtools}
\usepackage{enumitem}
\usepackage{braket}
\usepackage{xfrac}
\usepackage[capitalise]{cleveref}
\usepackage{amsfonts}

\def\d{\partial}
\def\nl{\nonumber \\}

\begin{document}
\title{Dark matter in the Randall-Sundrum model}

\author{Mukesh Kumar, Ashok Goyal, Rashidul Islam}

\address{School of Physics and Institute for Collider Particle Physics,
University of the Witwatersrand, Johannesburg, Wits 2050, South Africa.}

\ead{mukesh.kumar@cern.ch}

\begin{abstract}
In Ref.~\cite{Goyal:2019vsw}, we had considered simplified dark matter models interacting gravitationally with the Standard Model particles in a Randall-Sundrum frame work where models are considered with non-universal couplings. In the top-philic graviton model the right-handed top quarks are taken to interact strongly with the gravitons and in the lepton-philic model, we assume that only the right-handed charged leptons interact strongly with the gravitons. We extend the study to include not only the scalar, vector and spin-1/2 fermions but also spin-3/2 fermionic dark matter. We find that there is a large parameter space in these benchmark models where it is possible to achieve the observed relic density consistent with the direct and indirect searches, which are also consistent with the data from Large Hadron Collider.
\end{abstract}

\section{Introduction}
\label{sec:model}
The dark matter constitutes roughly 75\% of the entire matter existing in the Universe. The Planck collaboration has measured the DM density to great precision and has given the value of relic density $\Omega_{\rm DM} h^2 = 0.1198\pm0.0012$~\cite{Aghanim:2018eyx}. Though the nature of the DM particles remains elusive. In this work we consider bulk Randall-Sundrum (RS) model in which (i) only the right-handed top-quarks (top-philic) interact strongly with KK graviton~\cite{Lee:2014caa} or (ii) only the right-handed leptons (lepto-philic) interact strongly with the KK gravitons. We also extend the study to consider spin-3/2 DM particle, along with the scalar, vector and spin-1/2 fermionic DM particles.

\section{Dark matter in the Randall-Sundrum framework}
\label{sec:model}
In the RS framework, the particle interaction with the massive spin-2 gravitons $Y_{\mu\nu}$ is purely gravitational and is through the energy-momentum tensor being given by
\begin{align}
  {\cal L}_{\rm int} = - \sum \frac{c_i}{\Lambda} Y_{\mu\nu} T_i^{\mu\nu},
\end{align}
where $T_i^{\mu\nu}$ is the energy-momentum tensor of the $i^{\rm th}$ particle, $c_i$ is the corresponding coupling and $\Lambda$ is the scale of KK graviton interaction. We assume the DM fields to be either a real scalar, a real vector, a vector-like spin-1/2 Dirac fermion or a spin-3/2 fermion. The DM fields are further assumed to be SM singlets which do not carry any SM charge. It is also assumed that the DM particles are odd under a discrete ${\mathbb Z}_2$ symmetry so that there is no mixing between the DM and SM fields. It is assumed that the DM particles live on the IR brane. The mass of the DM particle is taken to be less than the mass of the gravitons for simplicity so that there is no DM annihilation into KK graviton states. We consider two benchmark models depending upon the relative placement of SM particles on or near the branes:
\begin{description}
  \item[Model A] In the first benchmark model (Top-philic KK graviton) the right-handed top-quarks alone are assumed to be located on the IR brane, $SU(3)_C$ and $U(1)_Y$ gauge bosons live in the bulk and the rest of the SM fields including the $SU(2)_L$ gauge bosons live on the UV brane or close to it. The interaction Lagrangian for this model is given as:
  \begin{align}
    {\cal L}_{\rm int}
    =&\,
    - \frac{1}{\Lambda}
    \Big[ i \frac{c_{tt}}{4} \left\{ \bar t_R \left(\gamma^\mu \overleftrightarrow{D}^\nu + \gamma^\nu \overleftrightarrow{D}^\mu \right) t
    - 2\, \eta^{\mu\nu} \bar t_R \overleftrightarrow{\slashed D} t_R \right\}
    \nl
    &\,
    + \frac{\alpha_S}{4\pi} c_{gg}
    \Big\{\frac{1}{4} \eta^{\mu\nu} G_a^{\lambda \rho} G^a_{\lambda \rho} - G_a^{\mu\lambda} G^{a\nu}_{\lambda} \Big\}
    + \frac{\alpha}{4\pi} c_1 \Big\{ \frac{1}{4}\eta^{\mu\nu} B^{\lambda\rho} B^{\lambda\rho} - B^{\mu\lambda} B^{\nu}_{\lambda} \Big\}
    \Big]\,Y_{\mu\nu}, \label{lintA}
  \end{align}  
where $D_\mu = \d_\mu + i (2/3)\,g_1 B_\mu + i g_s t^a G^a_\mu$ is the covariant derivative for the right-handed top quark field, $G_\mu$ and $B_\mu^a$ are the gluon and $U(1)_Y$ gauge boson fields respectively. The coupling $c_{tt}$ are scaled by appropriate loop suppression and $c_{tt} > c_{gg} \alpha_S/4\pi \sim c_1 \alpha/4\pi \gg$ other couplings.  
  
  \item[Model B] In the second benchmark model (lepto-philic KK graviton), the right-handed leptons are assumed to live on the IR brane and only the $U(1)_Y$ gauge bosons are assumed to live in the bulk with the rest of the SM particles including the gauge bosons live close to or on the UV brane and the interaction Lagrangian is as follows:
  \begin{align}
    {\cal L}_{\rm int}
    =&\,
    - \frac{1}{\Lambda}
    \Big[ i \sum_{\ell=e,\mu,\tau} \frac{c_{\ell\ell}}{4} \left\{ \bar\ell_R \left(\gamma^\mu \overleftrightarrow{D}^\nu + \gamma^\nu \overleftrightarrow{D}^\mu \right) \ell
    - 2\, \eta^{\mu\nu} \bar\ell_R \overleftrightarrow{\slashed D} \ell_R \right\}
    \nl
    &\,
    + \frac{\alpha}{4\pi} c_1 \Big\{ \frac{1}{4}\eta^{\mu\nu} B^{\lambda\rho} B^{\lambda\rho} - B^{\mu\lambda} B^{\nu}_{\lambda} \Big\}
    \Big]\,Y_{\mu\nu}, \label{lintB}
  \end{align}
  where $D_\mu = \d_\mu + i \frac{2}{3} g_1 B_\mu$ is the covariant derivative for the right-handed leptons.
\end{description}
The DM particles scalars ($S$), vectors ($V$), spin-1/2 fermions ($\chi$) and spin-3/2 fermions ($\Psi$) are taken to be present on the IR brane and interact with KK gravitons with a coupling strength of order one, through the energy-momentum tensor. The interaction Lagrangian is given as
\begin{align}
  {\cal L}_{\rm int} = -\frac{1}{\Lambda} c_{\rm DM} Y_{\mu\nu} T_{\rm DM}^{\mu\nu}.
\end{align}
For the expression of energy-momentum tensor $T_{\rm DM}^{\mu\nu}$ we refer~\cite{Goyal:2019vsw}. After electroweak symmetry breaking, the coupling between $Y_{\mu\nu}$ and $U(1)_Y$ gauge bosons is written in terms of coupling of $Y_{\mu\nu}$ with photons and $Z$ bosons as
\begin{gather}
  {\cal L}_{Y}
  \supset\,
 - \frac{1}{\Lambda} \Bigg[ \frac{\alpha}{4\pi} c_{\gamma\gamma} \left(\frac{1}{4}\eta^{\mu\nu} A^{\lambda\rho}A_{\lambda\rho} -  A^{\mu\lambda}A_{\lambda}^\nu \right) + \frac{\alpha}{4\pi} c_{Z\gamma} \left(\frac{1}{4}\eta^{\mu\nu} A^{\lambda\rho}Z_{\lambda\rho} -  A^{\mu\lambda}Z_{\lambda}^\nu \right)
  \nl
  + \frac{\alpha}{4\pi} c_{ZZ} \left(\frac{1}{4}\eta^{\mu\nu} Z^{\lambda\rho}Z_{\lambda\rho} -  Z^{\mu\lambda}Z_{\lambda}^\nu \right) \Bigg] Y_{\mu\nu}.
\end{gather}
The couplings of $c_{\gamma\gamma}, c_{\gamma Z}$ and $c_{ZZ}$ can be obtained from the coupling $c_1$ and are given as
\begin{gather}
\begin{rcases}
  c_{\gamma\gamma} = c_1 \cos^2\theta_W ,
  \\
  c_{Z\gamma} = - c_1 \sin 2\theta_W = - \sin 2\theta_W c_{\gamma\gamma}/ \cos^2\theta_W ,
  \\
  c_{ZZ} = c_1 \sin^2\theta_W = \tan^2\theta_W c_{\gamma\gamma} .
\end{rcases}
\end{gather}

Since the gravitons couple strongly with the right-handed top quarks in benchmark model {\bf A} and with right-handed charged leptons in model {\bf B}, the top quark and lepton triangle loop contribution to the graviton-gluon and graviton-$U(1)_Y$ gauge bosons can typically be of the same order as the corresponding tree-level couplings. The resulting effective loop induced couplings $c_{gg,\gamma\gamma}^{\rm eff}$ are evaluated in Refs.~\cite{Goyal:2019vsw,Geng:2016xin,Geng:2018hpq}.

\section{Dark matter phenomenology}
\label{sec:dmpheno}
\begin{figure*}[t]
  \centering
  \subfloat[]{\includegraphics[width=0.48\linewidth]{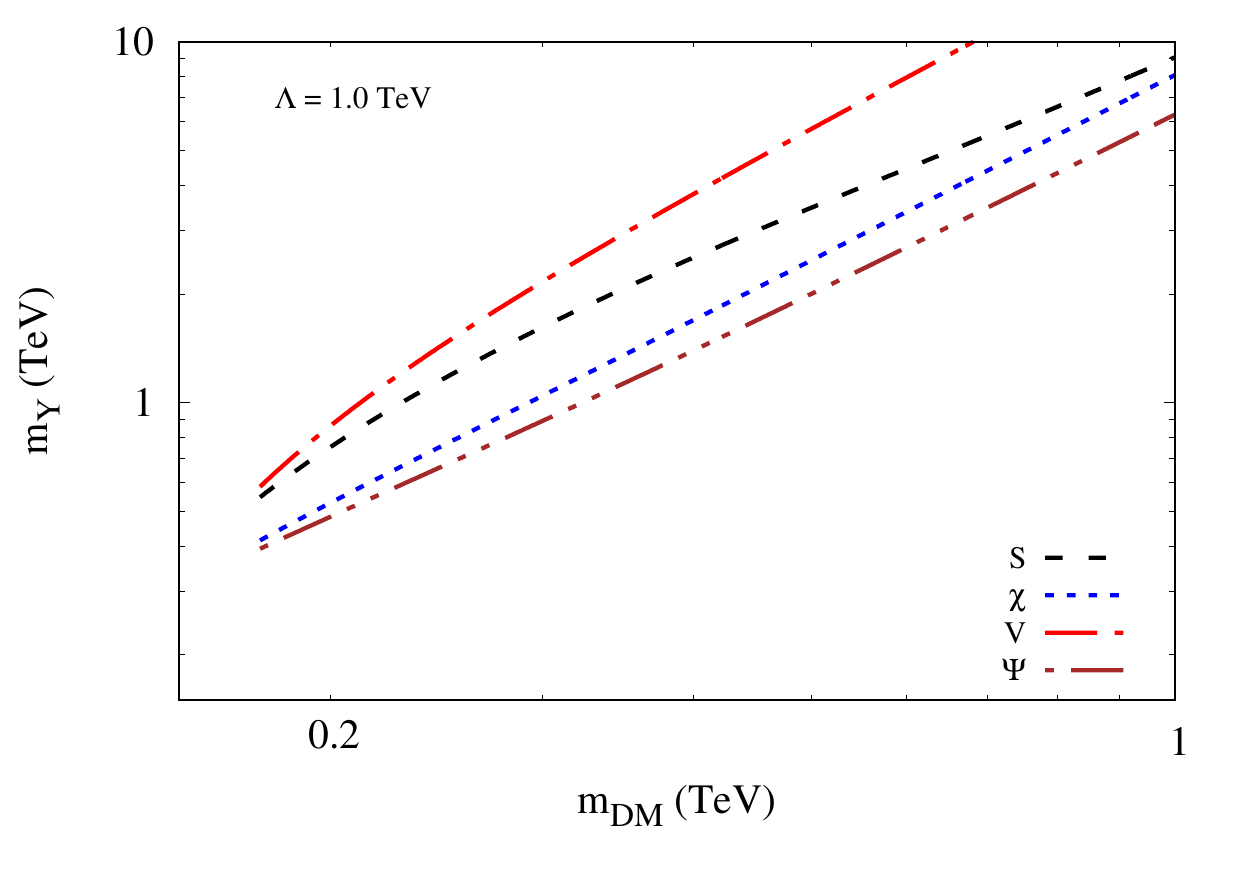}\label{fig:relicA1000}}
  \hfill
  \subfloat[]{\includegraphics[width=0.48\linewidth]{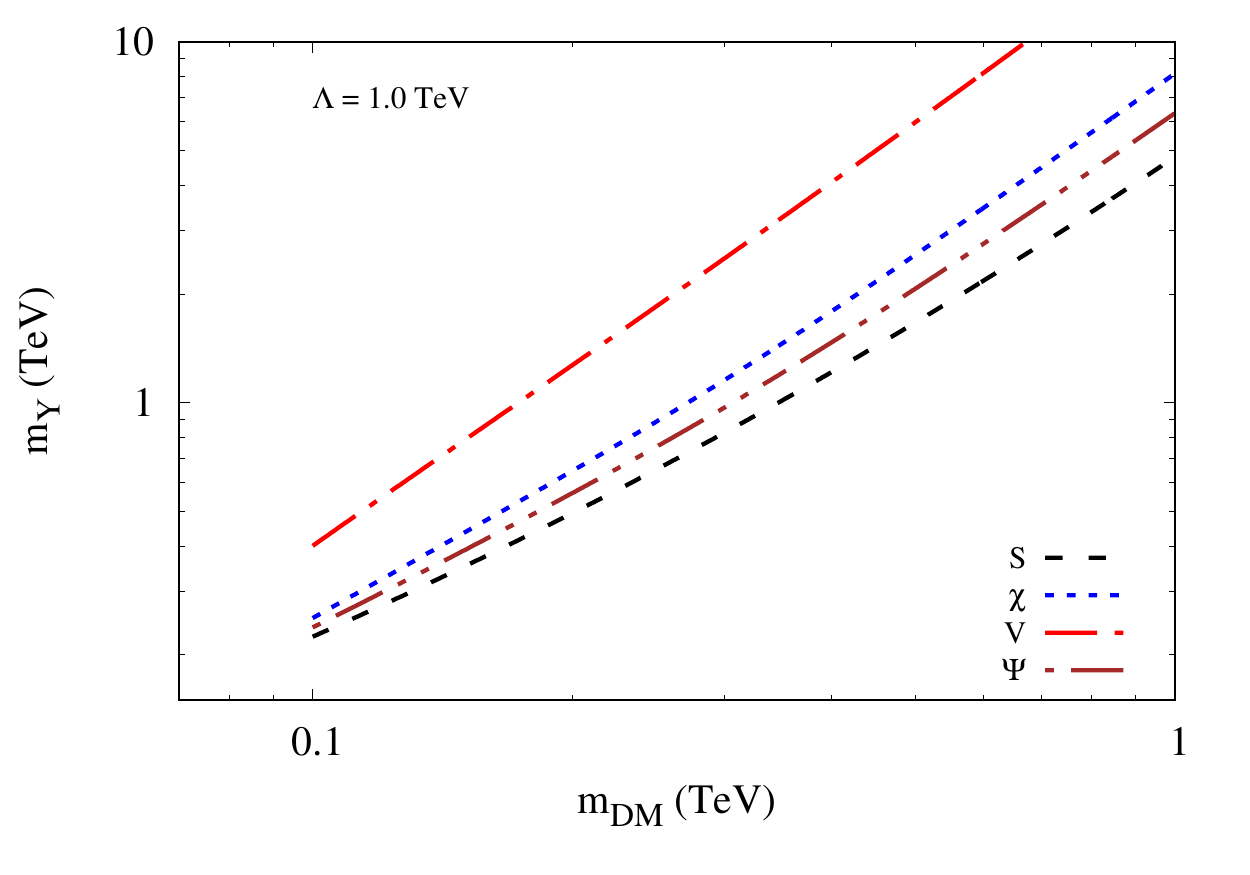}\label{fig:relicB1000}}
  \caption{Contours of constant relic density $\Omega_{\rm DM} h^2 = 0.119$ in the $m_{\rm DM} - m_Y$ plane for the case of scalar, spin-1/2, vector and spin-3/2 dark matter particles in the benchmark model (a) {\bf A} and (b) {\bf B} correspond to the KK-graviton interaction scale $\Lambda = 1$~TeV. The value of couplings $c_{tt}$, $c_{gg}$, $c_{ll}$ and $c_1$ are taken to be equal to~1.}
  \label{fig:relicAB}
\end{figure*}

\subsection{Thermal relic density}
\label{sec:relic}
The expansion of the Universe in the RS  model in the presence of radius stabilisation mechanism turns out to be in agreement with the effective 4-D description~\cite{Creminelli:2001th, Csaki:1999mp}. Thus in the early Universe, DM is in thermal equilibrium with the hot dense plasma and as the Universe expands and cools, it freezes out. The thermally averaged cross-section can be written as a one-dimensional integral over the centre of mass energy square $s$ as
\begin{align}
  \braket{\sigma | v}
  =
  \frac{1}{8\,m_{\rm DM}^2 \left[{\cal K}_2\left( m_{\rm DM}/T\right)\right]^2} \int\limits_{4 m_{\rm DM}^2}^\infty \left(s - 4 m_{\rm DM}^2 \right) \sqrt{s} \,\sigma_{\rm ann}\,{\cal K}_1 \left(\frac{\sqrt{s}}{T} \right) ds \,,
\end{align}
where ${\cal K}_1$ and ${\cal K}_2$ are the modified Bessel functions of the second kind, the annihilation cross-section $\sigma_{\rm ann}$ depends only on the masses, $s$ and couplings of the DM and SM particles involved.

The Boltzmann equation can be solved to give the thermal relic density
\begin{align}
 \Omega_{\rm DM} h^2
 \simeq
 \frac{1.07 \times 10^9\, x_F}{M_{\rm Pl} \sqrt{g^*(x_F)} \,\braket{\sigma | v}} \,.
\end{align}
Here $h$ is the Hubble parameter today, $g^*(x_F)$ is total number of dynamic degrees of freedom near the freeze-out temperature $T_F$ and $x_F = m_{\rm DM} / T_F$ is obtained by solving
\begin{align} 
  x_F
  =
  \ln \left[ c(c+2) \sqrt{\frac{15}{8}} \frac{g\,M_{\rm Pl}\,m_{\rm DM} \braket{\sigma | v}}{8 \pi^3\,\sqrt{x_F}\,\sqrt{g^*(x_F)} } \right]
\end{align}
and $c$ is of order~1. We have computed the relic density in the two benchmark models considered here numerically. The required model files generated using \textsc{LanHEP}~\cite{Semenov:2014rea} which calculates all the required couplings and Feynman rules by using the Lagrangian given in \cref{sec:model}. Then the model files are used in \textsc{CalcHEP}~\cite{Belyaev:2012qa} to calculate the required decay widths annihilation cross sections. In~\cref{fig:relicAB} we show the $2\sigma$ contour of constant relic density $0.119$ in the DM mass ($m_{\rm DM}$) and KK graviton mass ($m_Y$) for fixed KK graviton interaction scale $\Lambda$ for the case of scalar, spin-1/2, vector and spin-3/2 dark matter particles in the benchmark models {\bf A} and {\bf B} respectively.
\begin{figure*}[t]
  \centering
  \subfloat[]{\includegraphics[width=0.48\linewidth]{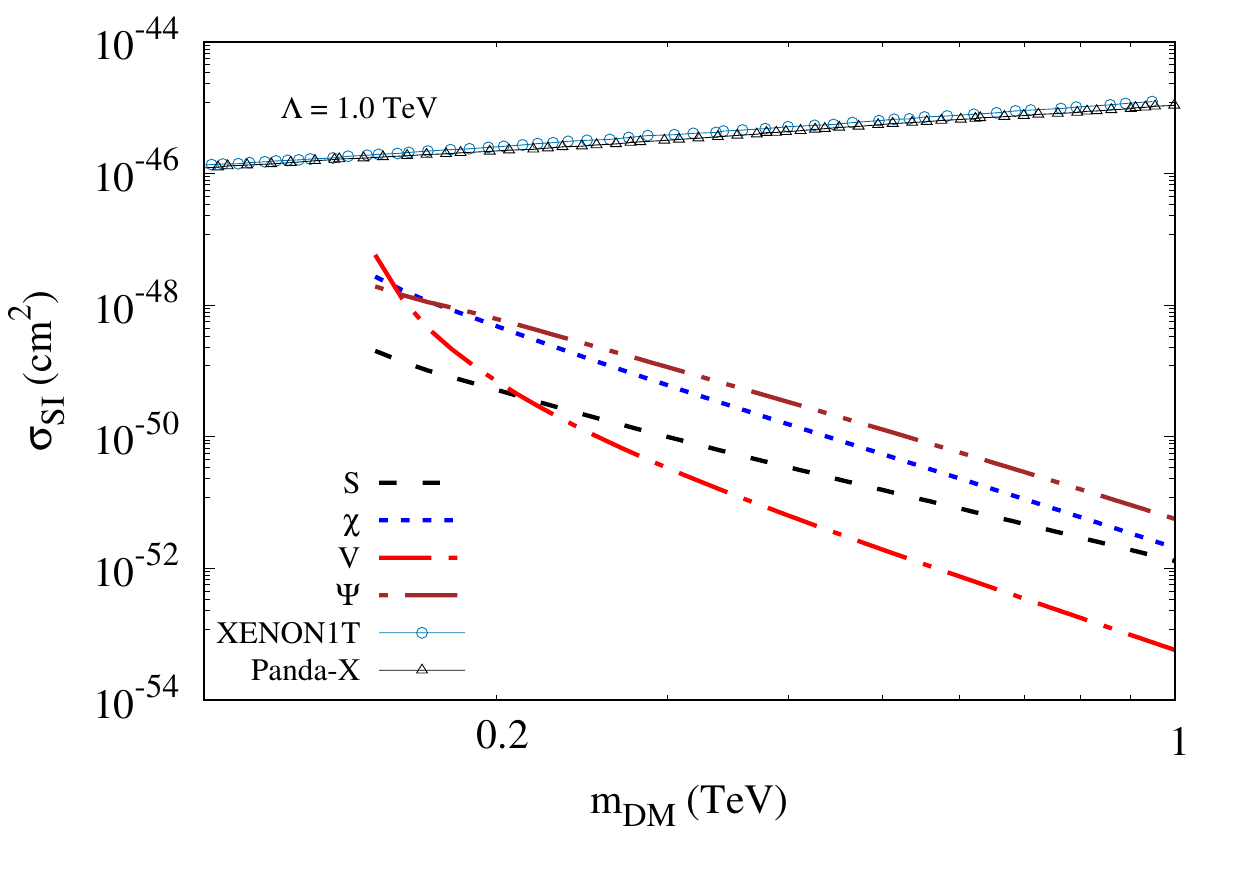}\label{fig:dd1000}}
  \hfill
  \subfloat[]{\includegraphics[width=0.48\linewidth]{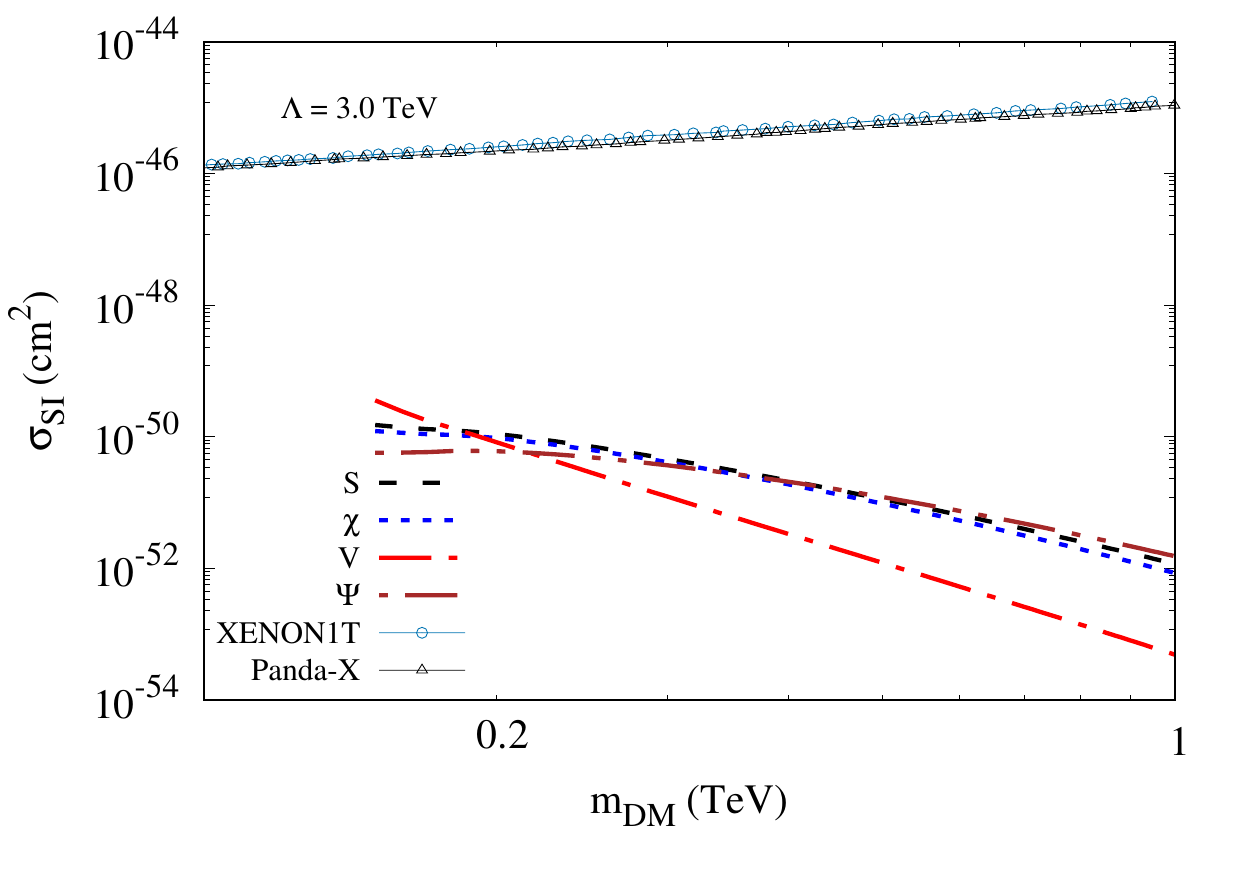}\label{fig:dd3000}}
  \caption{Dark matter-nucleon scattering cross-section as a function of DM mass in model {\bf A} correspond to the KK-graviton interaction scale (a) $\Lambda = 1$~TeV and (b) $\Lambda = 3$~TeV. Current bounds on spin-independent interactions from experiments like PANDA 2X-II 2017~\cite{Cui:2017nnn} and XENON1T~\cite{Aprile:2015uzo,Aprile:2017aty} are also shown.  All points on the contour are consistent with the observed relic density $\Omega_{\rm DM} h^2 = 0.119$.}
  \label{fig:dd}
\end{figure*}
\subsection{Direct detection}
\label{sec:dd}
In non-relativistic approximation, the spin-independent scattering cross-section does not depend on the spin of the DM particle. The spin-independent DM-nucleon scattering cross-section is given by
\begin{align}
  \sigma_{\rm SI}
  =&\,
  \frac{1}{729 \pi} \left(c_{gg}^{\rm eff}\right)^2 c_{\rm DM}^2 \left(\frac{\mu}{m_N}\right)^2 \left(\frac{m_N}{\Lambda\,m_Y} \right)^4 m^2_{\rm DM} \left[f_{TY}^{(N)}\right]^2
  \nl
  \simeq
  &\,
  1.75\times 10^{-49} \left(c_{gg}^{\rm eff} \right)^2 c_{\rm DM}^2  \left(\frac{m_{\rm DM}}{\rm TeV}\right)^2 \left(\frac{\rm TeV}{\Lambda}  \right)^4 \left(\frac{\rm TeV}{m_Y}  \right)^4\,{\rm cm^2}, {\label{sigmasi}}
\end{align}
where $m_N$ the nucleon mass and $\mu = {m_{\rm DM} m_N}/({m_{\rm DM} + m_N})$ is the reduced mass.

In the lepto-philic model {\bf B} considered here, DM-nucleon scattering arises through KK graviton-photon effective coupling which is suppressed by $\alpha/{4\pi}$. The DM-nucleon scattering cross-section is further suppressed by $\left(\alpha/{4\pi}\right)^2$ and does not give any meaningful constraint and is much below the sensitivity level achieved in the current or planned direct detection experiments~\cite{Kopp:2014tsa, Kopp:2009et, DEramo:2017zqw}. 
Using the expression~\cref{sigmasi} we have plotted the spin-independent DM-Nucleon scattering cross-section in the bench-mark model {\bf A}  as a function of DM mass in~\cref{fig:dd}. The parameter set used in the computation are consistent with the observed relic density given in~\cref{fig:relicAB}.  The upper limits from PANDA~2x-II 2017~\cite{Cui:2017nnn} and XENON-1T~\cite{Aprile:2015uzo,Aprile:2017aty} are also shown with the forbidden region.
\begin{figure*}[t]
  \centering
  \subfloat[]{\includegraphics[width=0.48\linewidth]{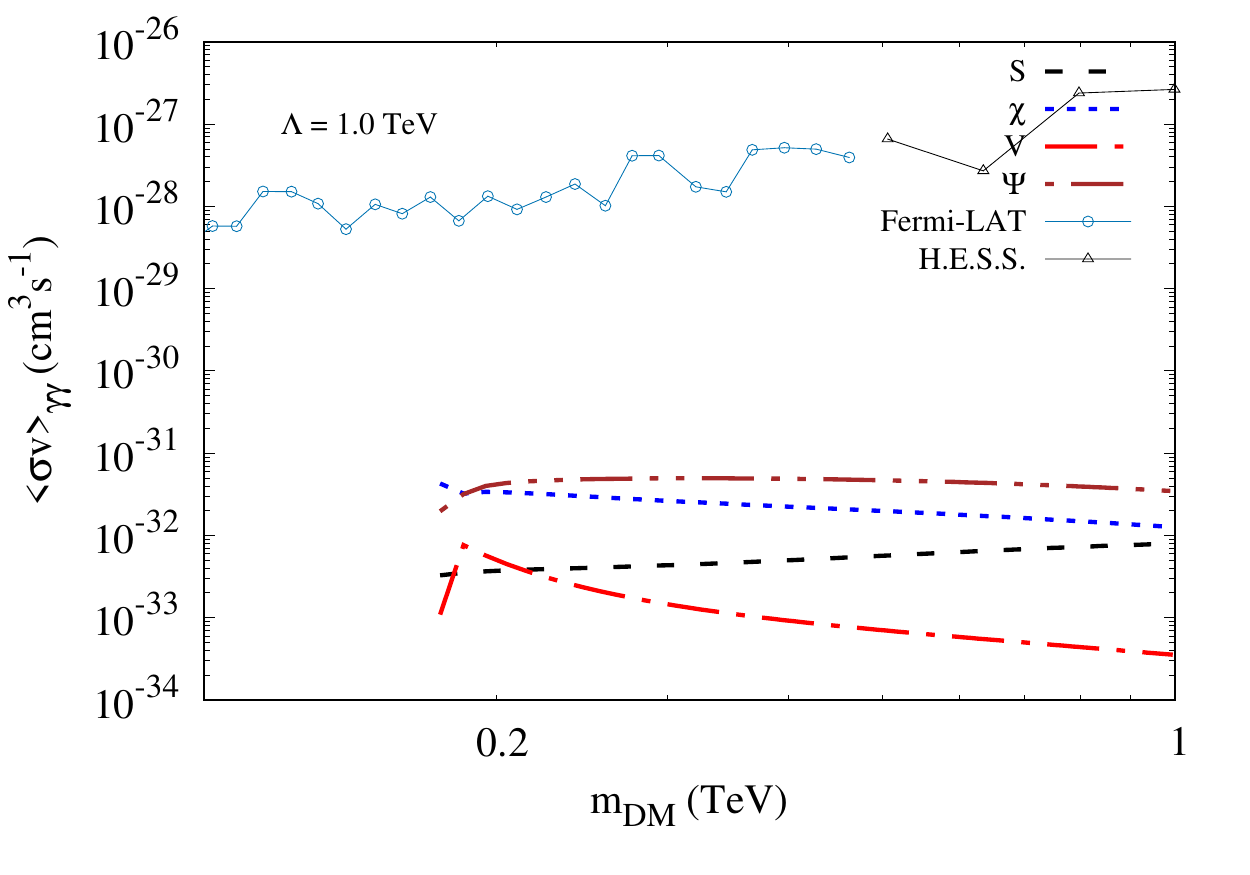}\label{fig:indirectA1000}}
  \hfill
  \subfloat[]{\includegraphics[width=0.48\linewidth]{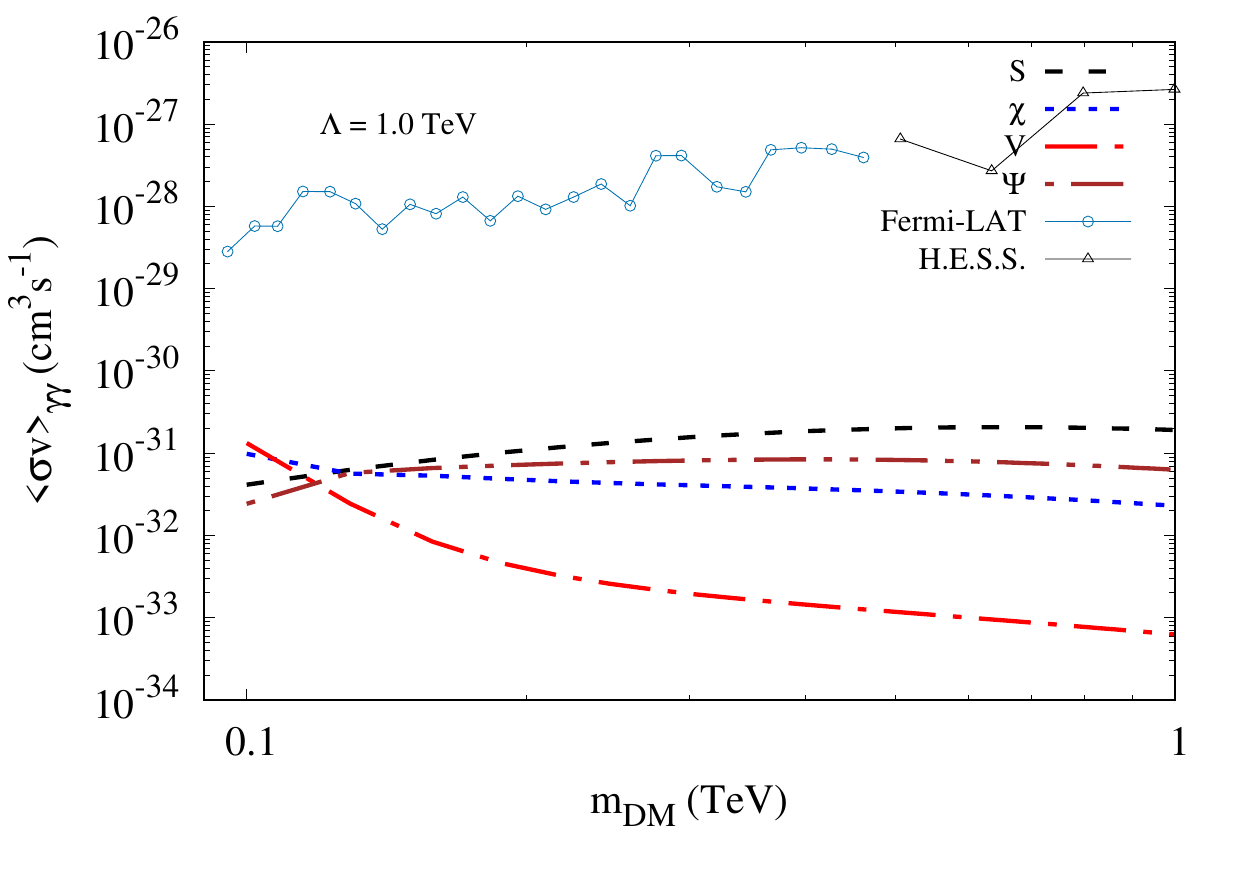}\label{fig:indirectB1000}}
  \caption{Velocity-averaged cross-section $\langle\sigma v\rangle_{\gamma \gamma}$ in the benchmark model (a) {\bf A} and (b) {\bf B} correspond to the KK-graviton interaction scale  $\Lambda = 1$~TeV. The current upper bounds from the Fermi-LAT~\cite{Ackermann:2015lka} and H.E.S.S.~\cite{Abramowski:2013ax,Abdalla:2016olq} data are shown. All points on the contour satisfy the observed relic density.}
  \label{fig:indirectAB}
\end{figure*}
\subsection{Indirect detection}
\label{sec:id}
Fermi-LAT~\cite{Ackermann:2015lka, Ackermann:2015zua} and H.E.S.S~\cite{Abramowski:2013ax,Abdalla:2016olq} have investigated DM annihilation as a possible source of incoming photon flux. These experiments are then used to put constraints on the upper limit of velocity averaged scattering cross-sections for various channels which can contribute to the observed photon flux. 
In~\cref{fig:indirectAB} we have plotted the variation of  velocity-averaged scattering cross-section $\langle\sigma v\rangle_{\gamma \gamma}$ with the DM mass in bench-mark model {\bf A} and {\bf B} respectively. We have also shown the observational constraints for the two photon final state annihilation rates. We find that the annihilation cross-sections for the above processes in the benchmark models {\bf A} and {\bf B} are roughly three to four orders of magnitude smaller than the current upper bounds from the Fermi-LAT~\cite{Ackermann:2015lka, Ackermann:2015zua} and H.E.S.S~\cite{Abramowski:2013ax,Abdalla:2016olq} data.

\begin{figure*}[t]
  \centering
  \subfloat[]{\includegraphics[width=0.48\linewidth]{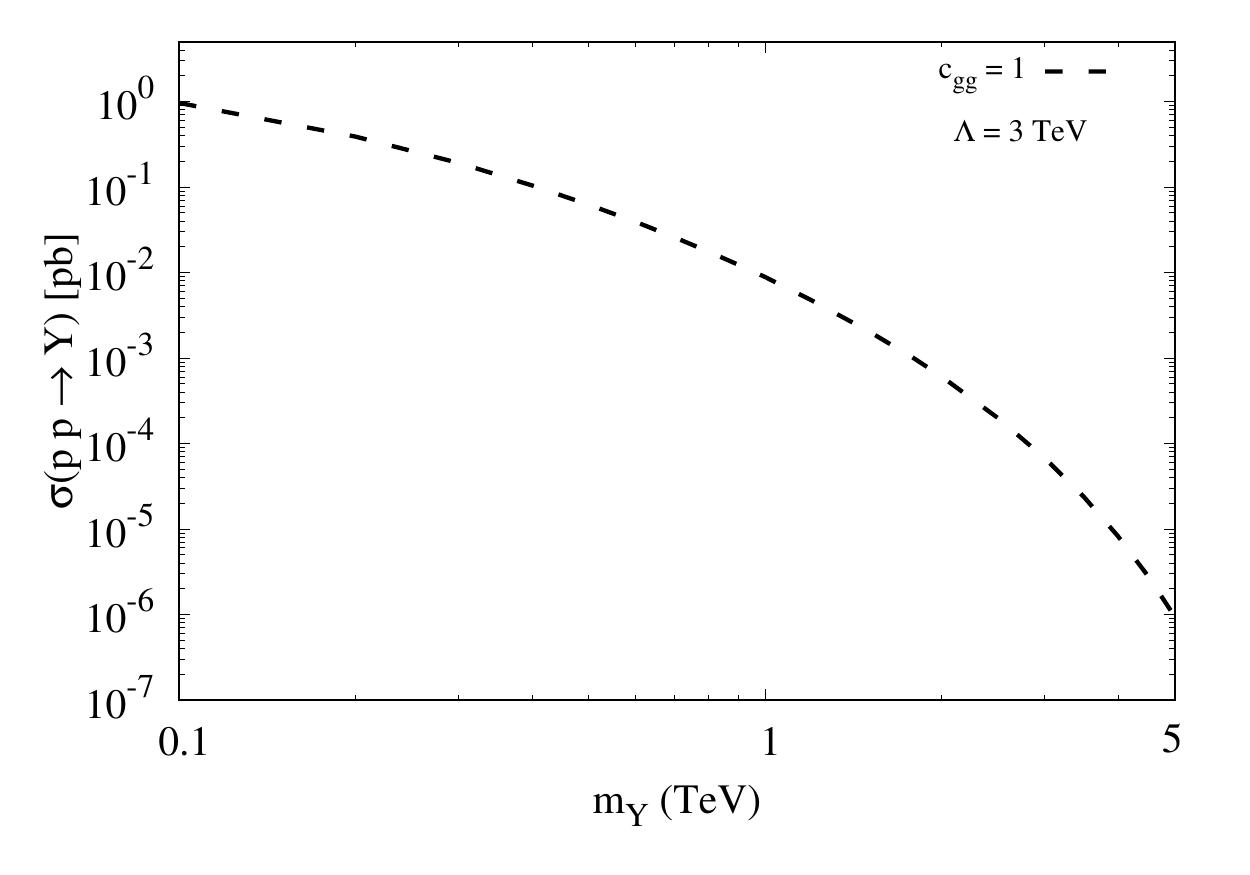}\label{fig:xsec}}
  \hfill
  \subfloat[]{\includegraphics[width=0.48\linewidth]{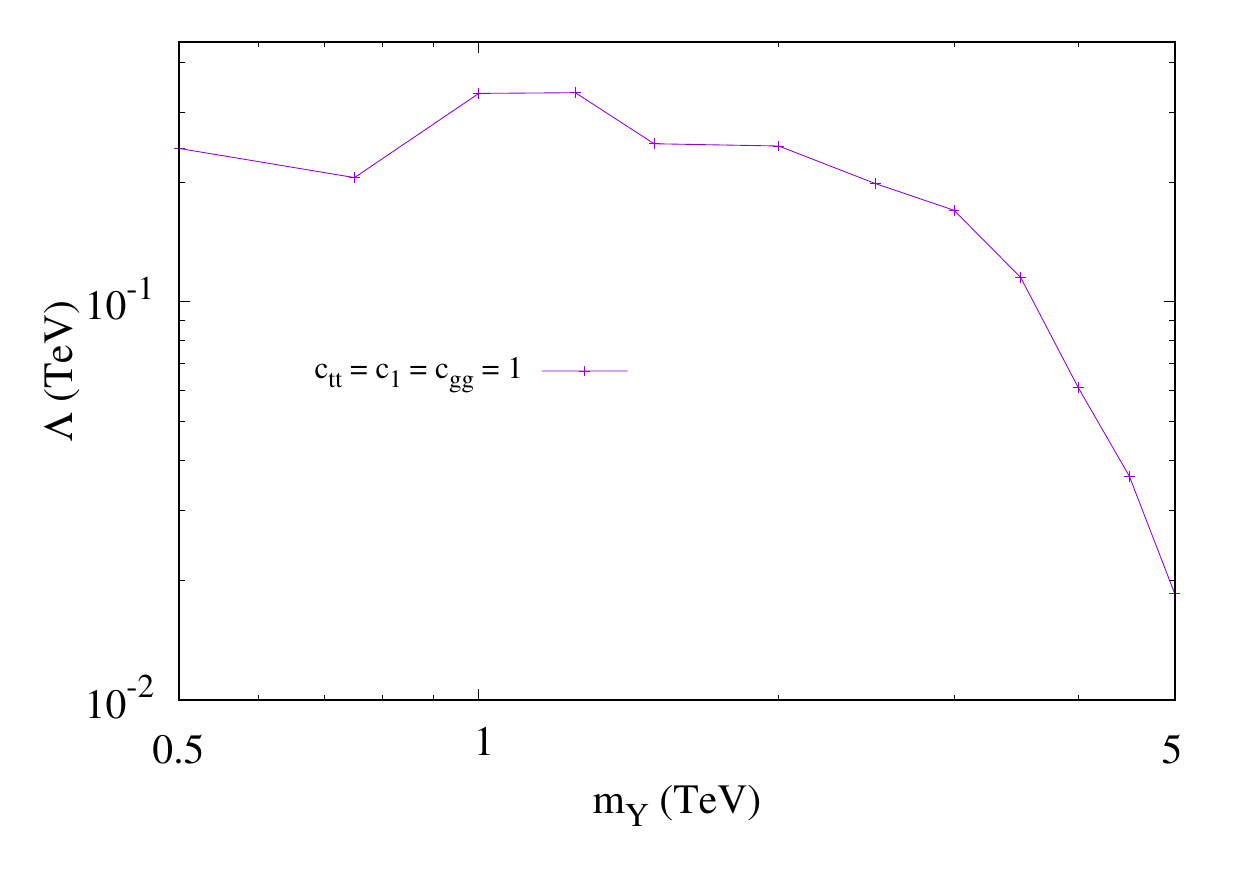}\label{fig:limit}}
  \caption{(a) KK graviton production LO cross-section multiplied by the K factor 1.3 through gluon fusion including the right-handed top loop. The cross-section is given for ${\sqrt s} = 13$~TeV for $\Lambda = 3$~TeV and the couplings $c_{tt} = c_{gg} =1$. The cross-section for different values of $\Lambda$ and the couplings can be obtained by scaling. (b) Constraints on the KK graviton interaction scale $\Lambda$ as a function of graviton mass from the observed 95\% C.L. lower limits of the resonance searches for ${\sqrt s} = 13$~TeV at 35.9~fb$^{-1}$ by CMS ~\cite{Sirunyan:2018ryr}.}
  \label{fig:xsecl}
\end{figure*}

\subsection{LHC constraints}
\label{sec:lhc}
The s-channel graviton ($Y$) production $p p \to Y$ at LHC in our benchmark model {\bf A} is dominated by gluon fusion and through the right-handed top quark loop. We calculate the $Y$ production leading order (LO) cross-section at ${\sqrt s} =13$~TeV at 35.9 fb$^{-1}$ . In~\cref{fig:xsec}, $Y$ production LO cross-section multiplied by the K-factor 1.3~\cite{Bonciani:2015hgv} is shown for the graviton mass $m_Y$ lying between 100~GeV to 5~TeV by using the effective coupling $c_{gg}^{\rm eff}$. The cross-section for a given value of $\Lambda$ can be obtained by scaling.

In order to estimate the constraints on the graviton interaction scale $\Lambda$, we employ the recent CMS~\cite{Sirunyan:2018ryr} data for resonance searches in the narrow-width approximation for $t \bar t$ production cross-section at ${\sqrt s} = 13$~TeV with a luminosity of 35.9~fb$^{-1}$ and taking the $Y \to t \bar t$ branching ratio to be one. The $t \bar t$ final state of the KK graviton gives the strongest constraint on the scale factor $\Lambda$. In~\cref{fig:limit}, we have shown the constraints on the graviton interaction scale $\Lambda$ from the observed 95\% C.L. lower limits given by CMS. We find that the lower limit on $\Lambda$ for all mass points $m_Y$ lying between 500~GeV to 5~TeV for which the data exists is less than a few hundred~GeV.

\section{Conclusions}
\label{sec:rc}
In conclusion, the top-philic as well as the lepto-philic models in the Randall-Sundram framework discussed in this work, are capable of explaining the observed relic density for a reasonable set of parameters without transgressing the constraints from the direct and indirect experiments. They are also consistent with the LHC data.

\section*{References}


\begin{thebibliography}{9}

\bibitem{Goyal:2019vsw} 
  A.~Goyal, R.~Islam and M.~Kumar,
  arXiv:1905.10583 [hep-ph].

\bibitem{Aghanim:2018eyx} 
  N.~Aghanim {\it et al.} [Planck Collaboration],
  arXiv:1807.06209 [astro-ph.CO].
  
\bibitem{Lee:2014caa} 
  H.~M.~Lee, M.~Park and V.~Sanz,
  JHEP {\bf 1405}, 063 (2014)
  [arXiv:1401.5301 [hep-ph]].
      
\bibitem{Geng:2016xin} 
  C.~Q.~Geng and D.~Huang,
  Phys.\ Rev.\ D {\bf 93}, no. 11, 115032 (2016)
  [arXiv:1601.07385 [hep-ph]].

\bibitem{Geng:2018hpq} 
  C.~Q.~Geng, D.~Huang and K.~Yamashita,
  JHEP {\bf 1810}, 046 (2018)
  [arXiv:1807.09643 [hep-ph]].
 
\bibitem{Creminelli:2001th} 
  P.~Creminelli, A.~Nicolis and R.~Rattazzi,
  JHEP {\bf 0203}, 051 (2002)
  [hep-th/0107141].

\bibitem{Csaki:1999mp} 
  C.~Csaki, M.~Graesser, L.~Randall and J.~Terning,
  Phys.\ Rev.\ D {\bf 62}, 045015 (2000)
  [hep-ph/9911406].

\bibitem{Semenov:2014rea} 
  A.~Semenov,
  Comput.\ Phys.\ Commun.\  {\bf 201}, 167 (2016)
  [arXiv:1412.5016 [physics.comp-ph]].

\bibitem{Belyaev:2012qa} 
  A.~Belyaev, N.~D.~Christensen and A.~Pukhov,
  Comput.\ Phys.\ Commun.\  {\bf 184}, 1729 (2013)
  [arXiv:1207.6082 [hep-ph]].

\bibitem{Kopp:2014tsa} 
  J.~Kopp, L.~Michaels and J.~Smirnov,
  JCAP {\bf 1404}, 022 (2014)
  [arXiv:1401.6457 [hep-ph]].
  
\bibitem{Kopp:2009et} 
  J.~Kopp, V.~Niro, T.~Schwetz and J.~Zupan,
  Phys.\ Rev.\ D {\bf 80}, 083502 (2009)
  [arXiv:0907.3159 [hep-ph]].

\bibitem{DEramo:2017zqw} 
  F.~D'Eramo, B.~J.~Kavanagh and P.~Panci,
  Phys.\ Lett.\ B {\bf 771}, 339 (2017)
  [arXiv:1702.00016 [hep-ph]].


\bibitem{Cui:2017nnn} 
  X.~Cui {\it et al.} [PandaX-II Collaboration],
  Phys.\ Rev.\ Lett.\  {\bf 119}, no. 18, 181302 (2017)
  [arXiv:1708.06917 [astro-ph.CO]].
  
\bibitem{Aprile:2015uzo} 
  E.~Aprile {\it et al.} [XENON Collaboration],
  JCAP {\bf 1604}, no. 04, 027 (2016)
  [arXiv:1512.07501 [physics.ins-det]].

\bibitem{Aprile:2017aty} 
  E.~Aprile {\it et al.} [XENON Collaboration],
  Eur.\ Phys.\ J.\ C {\bf 77}, no. 12, 881 (2017)
  [arXiv:1708.07051 [astro-ph.IM]].

\bibitem{Ackermann:2015lka} 
  M.~Ackermann {\it et al.} [Fermi-LAT Collaboration],
  Phys.\ Rev.\ D {\bf 91}, no. 12, 122002 (2015)
  [arXiv:1506.00013 [astro-ph.HE]].

\bibitem{Ackermann:2015zua} 
  M.~Ackermann {\it et al.} [Fermi-LAT Collaboration],
  Phys.\ Rev.\ Lett.\  {\bf 115}, no. 23, 231301 (2015)
  [arXiv:1503.02641 [astro-ph.HE]].
  
\bibitem{Abramowski:2013ax} 
  A.~Abramowski {\it et al.} [H.E.S.S. Collaboration],
  Phys.\ Rev.\ Lett.\  {\bf 110}, 041301 (2013)
  [arXiv:1301.1173 [astro-ph.HE]].
  
\bibitem{Abdalla:2016olq} 
  H.~Abdalla {\it et al.} [H.E.S.S. Collaboration],
  Phys.\ Rev.\ Lett.\  {\bf 117}, no. 15, 151302 (2016)
  [arXiv:1609.08091 [astro-ph.HE]].

\bibitem{Bonciani:2015hgv} 
  R.~Bonciani, T.~Jezo, M.~Klasen, F.~Lyonnet and I.~Schienbein,
  JHEP {\bf 1602}, 141 (2016)
  [arXiv:1511.08185 [hep-ph]].
  
\bibitem{Sirunyan:2018ryr} 
  A.~M.~Sirunyan {\it et al.} [CMS Collaboration],
  JHEP {\bf 1904}, 031 (2019)
  [arXiv:1810.05905 [hep-ex]].
  
  
  
\end{thebibliography}

\end{document}